\documentclass[prl,twocolumn,showpacs]{revtex4}
\usepackage{graphicx}
\usepackage{textcomp}
\usepackage{amsmath}

\begin{document}
\date{\today}
\author{D.~Schrader}
\author{I.~Dotsenko}
\author{M.~Khudaverdyan}
\author{Y.~Miroshnychenko}
\author{A.~Rauschenbeutel}
\author{D.~Meschede}
\email{meschede@iap.uni-bonn.de}

\affiliation{Institut f\"ur Angewandte Physik, Universit\"at Bonn,
Wegelerstr.~8, D-53115 Bonn, Germany}
\title{A neutral atom quantum register}

\begin{abstract}
We demonstrate the realization of a quantum register using a
string of single neutral atoms which are trapped in an optical
dipole trap. The atoms are selectively and coherently manipulated
in a magnetic field gradient using microwave radiation. Our
addressing scheme operates with a high spatial resolution and
qubit rotations on individual atoms are performed with 99~\%
contrast. In a final read-out operation we analyze each individual
atomic state. Finally, we have measured the coherence time and
identified the predominant dephasing mechanism for our register.
\end{abstract}
\date{\today}
\pacs{03.67.-a, 32.80.Pj, 39.25.+k, 42.50.Vk}

\maketitle

Information coded into the quantum states of physical systems
(qubits) can be processed according to the laws of quantum
mechanics. It has been shown that the quantum concepts of state
superposition and entanglement can lead to a dramatic speed up in
solving certain classes of computational
problems~\cite{Shor94,Grover97}. Over the past decade various
quantum computing schemes have been proposed. In a sequential
network of quantum logic gates quantum information is processed
using discrete one- and two-qubit operations~\cite{Lloyd95}.
Another approach is the one-way quantum computer which processes
information by performing one-qubit rotations and measurements on
an entangled cluster state~\cite{Raussendorf01}. All of these
schemes rely on the availability of a quantum register, i.~e.~a
well known number of qubits that can be individually addressed and
coherently manipulated. There are several physical systems, such
as trapped ions~\cite{Naegerl99,Schmidt-Kaler03b,Leibfried03},
nuclear spins in molecules~\cite{Vandersypen01}, or magnetic flux
qubits~\cite{Yamamoto03} that can serve as quantum registers.

Neutral atoms exhibit favourable properties for storing and
processing quantum information. Their hyperfine ground states are
readily prepared in pure quantum states including state
superpositions and can be well isolated from their environment. In
addition, using laser cooling techniques, countable numbers of
neutral atoms can be cooled, captured and
transported~\cite{Kuhr01,Schlosser01}. The coherence properties of
laser trapped atoms have been found to be adequate for storing
quantum information~\cite{Davidson95,Kuhr03}. Moreover, controlled
cold collisions~\cite{Mandel03} or the exchange of
microwave~\cite{Osnaghi01} or optical~\cite{Pellizzari95,You02}
photons in a resonator offer interesting schemes for mediating
coherent atom--atom interaction, essential for the realization of
quantum logic operations.

\begin{figure}
\begin{center}
  \includegraphics[width=\columnwidth]{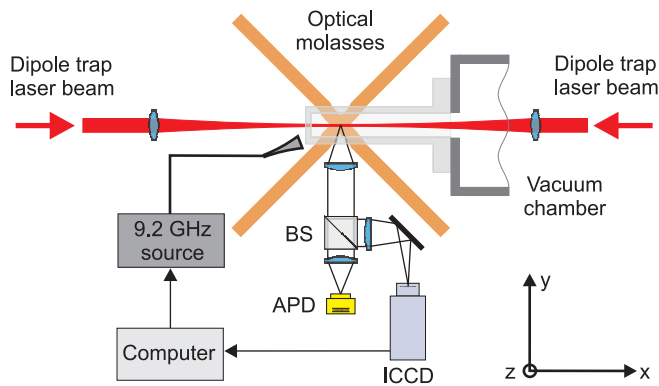}
\end{center}
\caption{Scheme of the experimental setup. Two focussed
counter-propagating Nd:YAG laser beams form the dipole trap. We
illuminate the trapped atoms by an optical molasses and split the
fluorescence light with a beamsplitter (BS) for imaging onto an
avalanche photo diode (APD) and an intensified CCD camera (ICCD).
Using the information about the atom positions, a computer
calculates the corresponding atomic resonance frequencies which
are then transmitted to the microwave source.}
\end{figure}

In our experiment we use a string of an exactly known number of
neutral caesium atoms. The atoms are trapped in the potential
wells of a spatially modulated, light induced potential created by
a far detuned standing wave dipole trap~\cite{Kuhr01,Schrader01}.
They can be optically resolved with an imaging system using an
intensified CCD camera (ICCD)~\cite{Alt02a,Miroshnychenko03}. Our
experimental setup is schematically depicted in Fig.~1. Two
focussed counter-propagating Nd:YAG laser beams at a wavelength of
$\lambda=1064$~nm generate the trapping potential with a depth of
up to 2.1~mK. This dipole trap is loaded from a high-gradient
magneto-optical trap (MOT). We determine the exact number of atoms
from the discrete levels of fluorescence of the
MOT~\cite{Haubrich96}. The transfer efficiency between the traps
is close to 100~$\%$. The storage time of 25~s in the dipole trap
is limited by collisions with the background gas. In order to
image the atoms, we illuminate them with a red-detuned
three-dimensional optical molasses which provides Doppler cooling.
The fluorescence light is observed by means of the ICCD with a
spatial resolution of 2.7~$\mu$m. Further details of the setup can
be found in previous
publications~\cite{Kuhr01,Kuhr03,Schrader01,Miroshnychenko03,Alt02b}.

\begin{figure}
\begin{center}
  \includegraphics[width=\columnwidth]{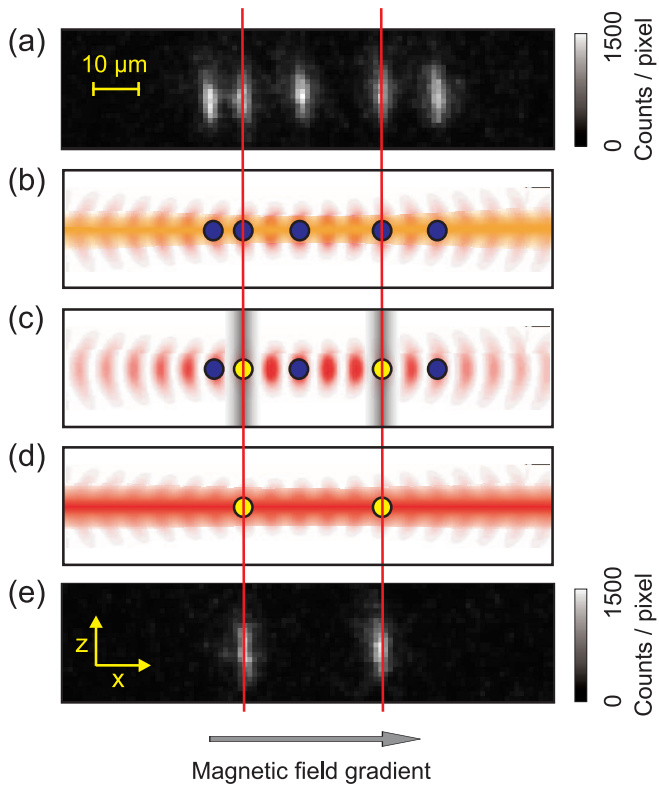}
\end{center}
\caption{5-atom quantum register. (a) Image of five neutral atoms
trapped in separate potential wells of a standing wave dipole
trap. The exposure time is 500~ms. One detected photon induces on
average 350 counts on the CCD chip. (b) An optical pumping laser
initializes the register in state $|00000\rangle$. (c) Two
microwave pulses at the resonance frequencies of atoms 2 and 4
perform a spin flip on these atoms to switch to state
$|01010\rangle$. The colours indicate the atomic states, blue
corresponding to state $|0\rangle$ and yellow to state
$|1\rangle$. (d) We state-selectively detect the atoms by applying
a push-out laser which removes atoms in state $|0\rangle$ from the
trap. (e) A final camera picture confirms the presence of atoms 2
and 4. Note that the spatial period of the schematic potential
wells in (b)--(d) is stretched for illustration purposes.}
\end{figure}

Fig.~2~(a) shows a picture of a string of five trapped atoms.
After their transfer from the MOT we let the atoms freely expand
along the dipole trap by switching off one of the dipole trap
laser beams for 1~ms. Following this expansion the atoms are
distributed over an interval of roughly 100~$\mu$m in the standing
wave trap.

In order to spectroscopically resolve the individual atoms in such
a string we apply an inhomogeneous magnetic field which introduces
a position dependent hyperfine transition frequency via the Zeeman
effect. For experimental simplicity this field is created by means
of the coils which also produce the magnetic quadrupole field for
the MOT. To achieve the maximum position sensitivity, we work with
the stretched $6S_{1/2}$ hyperfine ground states,
$|F=4,m_F=4\rangle$ and $|F=3,m_F=3\rangle$, with the quantization
axis oriented along the dipole trap axis. These two levels serve
as the qubit states in our quantum register and will be denoted
$|0\rangle$ and $|1\rangle$, respectively. Our applied magnetic
field has the form
\begin{equation}\label{eq:B-field}
  \vec B(\vec r)=(B_x,B_y,B_z)= (B_0,0,0)- B^\prime\cdot (x,y,-2z)\ .
\end{equation}
A homogeneous offset field $B_0=4$~G shifts the $|0\rangle
\leftrightarrow |1\rangle$ transition frequency by
$\nu_0=-9.8$~MHz with respect to the unperturbed value at 9.2~GHz.
A gradient field $B^\prime \approx 15$~G/cm along the dipole trap
yields a position-dependent frequency shift of
$\nu^\prime=-3.69\pm0.04$~kHz/$\mu$m, determined in an initial
calibration measurement.

We determine the positions of the atoms along the trap axis by
analyzing an ICCD image of the atom string with a fitting routine.
From these positions the corresponding atomic resonance
frequencies are calculated and sent to the microwave generator.
This entire procedure takes about 1~s. We then initialize the
register in state $|00000\rangle \equiv |0\rangle_1 |0\rangle_2
|0\rangle_3 |0\rangle_4 |0\rangle_5$, where the subscript denotes
the atom number. For this purpose we switch on the magnetic field
and optically pump all atoms into state $|0\rangle$ with a
$\sigma^+$-polarized laser on the $F=4\leftrightarrow F'=4$
transition and a repumping laser on the $F=3\leftrightarrow F'=4$
transition of the $D2$ line, see Fig.~2~(b).

We now carry out single qubit operations on the initialized
register. In this demonstration we switch the register state from
$|00000\rangle$ to $|01010\rangle$. For this purpose, we perform
spin flips on atoms 2 and 4 by the sequential application of two
$\pi$-pulses at their respective frequencies, see Fig.~2~(c). To
measure the state of each qubit we switch off the magnetic field
and remove all atoms in state $|0\rangle$ from the trap by a
state-selective ``push-out'' laser~\cite{Kuhr03}, see Fig.~2~(d).
This detection scheme has an efficiency of better than 99~\%,
i.~e.~less than 1~\% of all atoms in state $|1\rangle$
($|0\rangle$) are erroneously detected in state $|0\rangle$
($|1\rangle$). The presence or absence of each atom in the
subsequently taken image therefore reveals its state, $|1\rangle$
or $|0\rangle$, respectively. As expected, atoms 2 and 4 are
present in Fig.~2~(e), while atoms 1, 3, and 5 have been removed
from the trap.

In order to characterize the performance of our scheme we
determine its resolution, i.~e.~the minimum distance between
adjacent atoms necessary for selective addressing. For this
purpose, we trap only one atom at a time in our dipole trap and
initialize it in state $|0\rangle$. Then we apply a $\pi$-pulse to
the atom, with a Gaussian shaped microwave amplitude $A(t)=A_0
\exp{(-t^2/2\sigma_{\tau}^2)}$. The frequency of this microwave
pulse is detuned from the atomic resonance frequency at the
position of the atom. We record the population transfer from
$|0\rangle$ to $|1\rangle$ as a function of this detuning $\delta$
which corresponds to a position offset $\Delta
x=\delta/\nu^\prime$. For this purpose, we subject the atom to the
state-selective push-out laser and reveal its presence or absence
through fluorescence detection after retransferring it to the MOT.

\begin{figure}
\begin{center}
  \includegraphics[width=\columnwidth]{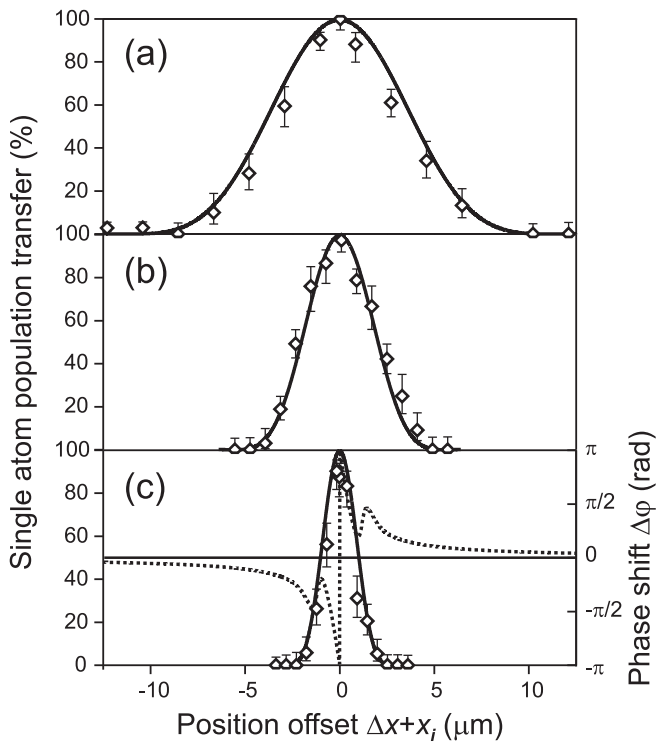}
\end{center}
\caption{Measurement of the addressing resolution. The data points
show the population transfer efficiency of one atom being exposed
to a microwave $\pi$-pulse resonant with a position $\Delta x$
away from the calculated atom position. Each point consists of
approximately 40 single atom events. Resonant addressing reveals a
spin flip efficiency of close to 100~$\%$. For longer pulses
($2\sigma_{\tau}=17.7~\mu$s (a), 35.4~$\mu$s (b), and 70.7~$\mu$s
(c)) the spectra become narrower, with an addressing resolution of
up to $\sim 2.5~\mu$m (c). The center of the spectrum is slightly
shifted by $x_i$ due to slow drifts of the atomic resonance
frequency that occurred during the 10-hour data acquisition time.
Here, $x_a=-1.0~\mu$m, $x_b=-2.1~\mu$m, and $x_c=-3.4~\mu$m. The
measured data are in good agreement with a numerical simulation
(solid lines). For the parameters in (c) we show the relative
phase shift between states $|0\rangle$ and $|1\rangle$, obtained
from the same simulation (dotted line).}
\end{figure}

The result of this measurement is shown in Fig.~3~(a)--(c) for
different durations of the microwave pulse. Due to the narrowing
Fourier spectrum of the corresponding $\pi$-pulses, the spatial
interval of significant population transfer decreases with
increasing pulse duration. A pulse of length
$2\sigma_{\tau}=70.7~\mu$s, see Fig.~3~(c), swaps the state of an
atom at one position while an atom trapped at a site $2.5~\mu$m
away remains in its initial state with a probability of
$100^{+0}_{-2.7}~\%$. The predominant limitation for the
addressing resolution are slow drifts of the intensity and the
polarization of the dipole trap laser beams which change the
atomic resonance frequency by up to 1~kHz/h.

The maximum population transfer for resonant addressing is
$98.7^{+1.1}_{-3.0}~\%$ for a pulse length of up to $35~\mu$s.
This efficiency includes all experimental imperfections: losses
during transfer of the atom between the two traps and during
illumination of the atom in the dipole trap, imperfect state
initialization by optical pumping, and erroneous detection of the
atomic state. Fig.~3 also shows that the measured spectra are in
very good agreement with the theoretical prediction from a
numerical Bloch-vector simulation with no adjusted parameters. The
same simulation also allows us to calculate the coherent shift
$\Delta \varphi$ of the relative phase between states $|0\rangle$
and $|1\rangle$ induced in adjacent atoms due to non-resonant
interaction with the detuned microwave pulse. For the experimental
parameters of Fig.~3~(c) and an atom separation of
e.~g.~2.5~$\mu$m, this calculation yields a phase shift $\Delta
\varphi=0.2\cdot\pi$. It can be taken into account in further gate
operations.

\begin{figure}
\begin{center}
  \includegraphics[width=\columnwidth]{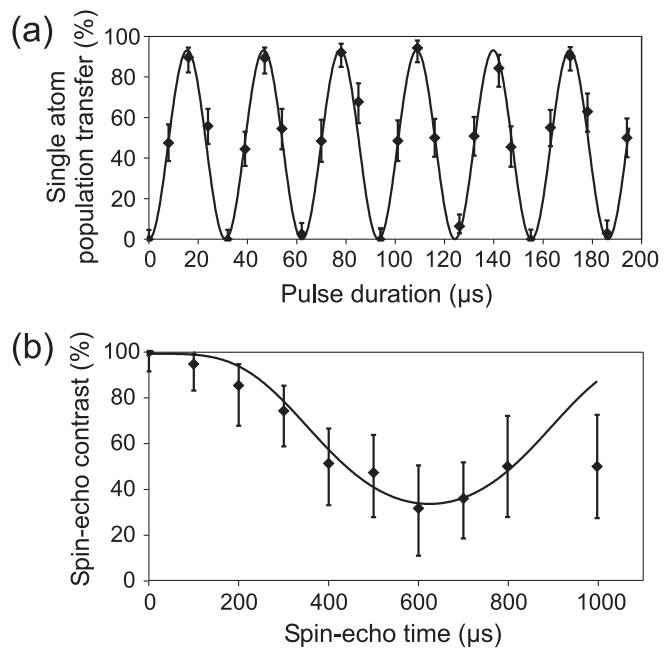}
\end{center}
\caption{Coherent qubit manipulation. (a) Demonstration of qubit
rotations of individually addressed single atoms. Each point shows
the averaged population transfer of approximately 40 single atom
events. The Rabi-oscillations have a contrast of
$99.1^{+0.9}_{-3.7}\%$. (b) Contrast of the spin-echo signal of
individually addressed atoms as a function of the spin-echo time.
The decrease of the spin-echo contrast to 35~\% after 600~$\mu$s
is caused by radial oscillations of the atoms in an inhomogeneous
magnetic field. The fit from a theoretical model (solid line) is
in good agreement with the measured data.}
\end{figure}

Arbitrary qubit rotations of our quantum register are demonstrated
in Fig.~4~(a). Here, we trap, image, and initialize one atom qubit
as above. After the application of a square microwave pulse of
duration $t_{\rm pulse}$ at the corresponding resonance frequency
of the atom, we measure the transfer probability to state
$|1\rangle$. Every point in Fig.~4~(a) corresponds to the transfer
probability deduced from approximately 40 single atom events. The
signal shows Rabi oscillations of the population between states
$|0\rangle$ and $|1\rangle$. This state evolution reads
$|\psi(t_{\rm pulse})\rangle=\cos(\Omega_R t_{\rm pulse}/2)
|0\rangle - i\sin(\Omega_R t_{\rm pulse}/2)|1\rangle$, where
$\Omega_R=2\pi\cdot32$~kHz is the Rabi frequency. A pulse duration
of $\tau=\pi/2\Omega_R=8~\mu$s therefore corresponds to a one
qubit Hadamard-gate in quantum information processing. The line in
Fig.~4~(a) is a sinusoidal fit which yields a contrast of
$99.1^{+0.9}_{-3.7}\%$. We hereby demonstrate reliable single
qubit rotations on our quantum register.

In order to investigate the coherence properties of the quantum
register we have performed a spin-echo measurement~\cite{Kuhr03}
on single atoms addressed in the magnetic field gradient. The
spin-echo contrast is shown in Fig.~4~(b) as a function of the
spin-echo time. We measure a reduction of the contrast to
approximately 35~\% within 600~$\mu$s. Note that this dephasing
time is two orders of magnitude larger than the measured one-qubit
switching time. A possible two-qubit gate performed by the
exchange of single photons in an optical high-finesse cavity has
an expected operation time of 200~ns and would be more than three
orders of magnitude faster than the dephasing time of our quantum
register~\cite{You02}.

The solid line in Fig.~4~(b) is a theoretical fit which models the
effect of the thermal oscillations of the trapped atoms inside the
applied inhomogeneous magnetic field. According to
equation~(\ref{eq:B-field}) we see that the modulus of the
magnetic field in radial direction varies as
\begin{equation}\label{eq:B-field_abs}
    |\vec{B}(x=0,y,z)|\approx B_0 + \frac{1}{2B_0}
    {B^\prime}^2(4z^2+y^2)\ .
\end{equation}
The Zeeman shift of the $|1\rangle \leftrightarrow |0\rangle$
transition therefore depends on the radial position of the atoms.
Consequently, a radial oscillation will result in a time varying
atomic resonance frequency and cause a dephasing between states
$|0\rangle$ and $|1\rangle$. Due to the oscillatory behaviour of
this dephasing, the resulting reduction of the spin-echo contrast
varies periodically. Complete rephasing is theoretically possible
for a spin-echo time of twice the radial oscillation period.
However, the maximum time interval in which quantum operations can
continuously be performed is determined by the initial decay of
the spin-echo contrast.

We find very good agreement between our model and the experimental
data for typical experimental parameters (see Fig.~4~(b)): a
temperature of the atomic ensemble of 80~$\mu$K, a radial
oscillation frequency of 1.6~kHz, and the dipole trap axis running
15~$\mu$m above or below the symmetry plane of the B-field due to
alignment imperfections. This result indicates that the initial
reduction of our spin-echo contrast is predominantly caused by the
thermal radial oscillations of the atoms in our trap. Possible
ways to extend the coherence time of our quantum register
therefore include an increase of the magnetic offset field $B_0$
(see Eq.~(\ref{eq:B-field_abs})), a more advantageous magnetic
field geometry, a reduction of the radial oscillations by further
cooling of the atoms, and a transfer of the atoms to
decoherence-free states between addressing operations.

Summarizing our results, we have demonstrated that a string of
caesium atoms trapped in our standing wave dipole trap can be used
as a quantum register. We have initialized, selectively addressed,
coherently manipulated, and state-selectively detected the
hyperfine states of individual atoms within the string. Our scheme
operates on atoms separated by distances as small as 2.5~$\mu$m.
Therefore, if the qubits were evenly spaced, populating every
fifth trapping site, it is scalable to a few hundred qubits,
limited by the axial extension of the trapping potential.

Currently, we set up a second, perpendicular conveyor-belt which
should allow us to place each individual atom of the quantum
register into a desired potential well. This would enable us to
distribute the atoms evenly in the trapping region. Furthermore,
it should permit to induce controlled interaction of arbitrary
pairs of distant atom qubits by placing them next to each other.
Our scheme is compatible with the requirements of cavity quantum
electrodynamics or controlled cold collision experiments, making
our quantum register a versatile tool for the implementation of
quantum logic operations.

We thank W.~Alt and S.~Kuhr for valuable discussions and technical
assistance. This work has been supported by the Deutsche
Forschungsgemeinschaft and the European Commission.


\begin{thebibliography}{99}
\bibitem{Shor94}
P.~Shor, SIAM J. Comp. {\bf 26}, 1484 (1997).

\bibitem{Grover97}
L.~K.~Grover, Phys. Rev. Lett. {\bf 79}, 325 (1997).

\bibitem{Lloyd95}
S.~Lloyd, Phys. Rev. Lett. {\bf 75}, 346 (1995).

\bibitem{Raussendorf01}
R.~Raussendorf and H.~J. Briegel, Phys. Rev. Lett. {\bf 86}, 5188
(2001).

\bibitem{Naegerl99}
H.~C. N\"agerl {\it et~al.\/}, Phys. Rev. A {\bf 60}, 145 (1999).

\bibitem{Schmidt-Kaler03b}
F.~Schmidt-Kaler {\it et~al.\/}, Nature {\bf 422}, 408 (2003).

\bibitem{Leibfried03}
D.~Leibfried {\it et~al.\/}, Nature {\bf 422}, 412 (2003).

\bibitem{Vandersypen01}
L.~M.~K. Vandersypen {\it et~al.\/}, Nature {\bf 414}, 883 (2001).

\bibitem{Yamamoto03}
T.~Yamamoto {\it et~al.\/}, Nature {\bf 425}, 941 (2003).

\bibitem{Kuhr01}
S.~Kuhr {\it et~al.\/}, Science {\bf 293}, 278 (2001). [published
  online; 10.1126/science.1062725].

\bibitem{Schlosser01}
N.~Schlosser, G.~Reymond, I.~Protsenko, and P.~Grangier, Nature
{\bf 411}, 1024 (2001).

\bibitem{Davidson95}
N.~Davidson {\it et~al.\/}, Phys. Rev. Lett. {\bf 74}, 1311
(1995).

\bibitem{Kuhr03}
S.~Kuhr {\it et~al.\/}, Phys. Rev. Lett. {\bf 91}, 213002 (2003).

\bibitem{Mandel03}
O.~Mandel {\it et~al.\/}, Nature {\bf 425}, 937 (2003).

\bibitem{Osnaghi01}
S.~Osnaghi {\it et~al.\/}, Phys. Rev. Lett. {\bf 87}, 037902
(2001).

\bibitem{Pellizzari95}
T.~Pellizzari, S.~A. Gardiner, J.~I. Cirac, and P.~Zoller, Phys.
Rev. Lett. {\bf 75}, 3788 (1995).

\bibitem{You02}
L.~You, X.~X. Yi, and X.~H. Su, Phys. Rev. A {\bf 67}, 032308
(2003).

\bibitem{Schrader01}
D.~Schrader {\it et~al.\/}, Appl. Phys. B {\bf 73}, 819 (2001).

\bibitem{Alt02a}
W.~Alt, Optik {\bf 113}, 142 (2002).

\bibitem{Miroshnychenko03}
Y.~Miroshnychenko {\it et~al.\/}, Optics Express {\bf 11}, 3498
  (2003).

\bibitem{Haubrich96}
D.~Haubrich {\it et~al.\/}, Europhys. Lett. {\bf 34}, 663 (1996).

\bibitem{Alt02b}
W.~Alt {\it et~al.\/}, Phys. Rev. A {\bf 67}, 033403 (2003).

\end{thebibliography}
\end{document}